\documentclass[a4paper]{jpconf}
\usepackage{graphicx}
\usepackage{amsmath}
\usepackage{amssymb}
\usepackage{textcomp}
\usepackage[square,sort&compress,numbers]{natbib}
\usepackage[T1]{fontenc}
\usepackage[latin1]{inputenc}
\newcommand{\vr}{\mathbf{r}}
\newcommand{\vR}{\mathbf{R}}

\begin{document}
\title{Magnetic-optical transitions induced by twisted light in quantum dots}

\author{G. F. Quinteiro$^1$, D. E. Reiter$^2$, T. Kuhn$^2$}

\address{$^1$Departamento de F\'{i}sica and IFIBA, FCEN, Universidad de Buenos Aires,
Buenos Aires, Argentina\\
$^2$Institut f\"ur Festk\"orpertheorie, Universit\"at M\"unster, 48149 M\"unster, Germany}

\ead{gquinteiro@df.uba.ar}

\begin{abstract}
It has been theoretically predicted that light carrying orbital angular
momentum, or twisted light, can be tuned to have a strong magnetic-field
component at optical frequencies. We here consider the interaction of these
peculiar fields with a semiconductor quantum dot and show that the magnetic
interaction results in new types of optical transitions. In particular, a
single pulse of such twisted light can drive light-hole-to-conduction band
transitions that are cumbersome to produce using conventional Gaussian beams
or even twisted light with dominant electric fields.
\end{abstract}

\noindent\textbf{Introduction}\\
Twisted light (TL) is light having a helical wave front. Such light exhibits
several interesting features: Due to the azimuthal phase dependence a phase
singularity occurs at the beam axis, leading to the name optical vortex.
Furthermore, in addition to spin angular momentum (SAM) associated with the
handedness of circular polarization, such light fields carry orbital angular
momentum (OAM). Depending on the combination of SAM and OAM, TL beams may
have strong field components along the propagation direction or strong
magnetic field close to the beam axis. The research in TL spans nowadays
several areas of fundamental and applied physics \cite{andrews2011str}
including its potential for quantum communication
\cite{mair2001entanglement}.

Of particular interest is the interaction of a TL beam with matter, which
opens up the possibility to address unusual transitions
\cite{quinteiro2015formulation,quinteiro2017formulation} and to excite
typically dark modes in plasmonics \cite{kerber2017reading}. In this paper we
focus on the interaction of TL with a semiconductor quantum dot (QD). QDs are
discussed for many applications in optoelectronics and spintronics
\cite{chen2004theory} which, however, requires a precise control of the
optically excited states. Here we will show that the interaction of a QD with
TL having a strong magnetic component may be used for the excitation of
specific light-hole (LH) excitons, which cannot easily be addressed by
conventional Gaussian beams or even twisted light with dominant electric
fields.\\[0.5ex]

\noindent\textbf{Twisted light with a strong magnetic component}\\
TL can be characterized by the handedness of its circular polarization
$\sigma$ (SAM) and its OAM quantum number $\ell$. We recently identified two
distinct classes of TL
\cite{quinteiro2015formulation,quinteiro2017formulation}, that we named
parallel and antiparallel class reflecting the relative orientation of SAM
and OAM. Figure 1 depicts the electric field profiles for a beam with
$\ell=2$ and $\sigma = \mp 1$ showing the pronounced difference in the
topology of the field lines.

\begin{figure}[h!]
\includegraphics[width=0.55\textwidth]{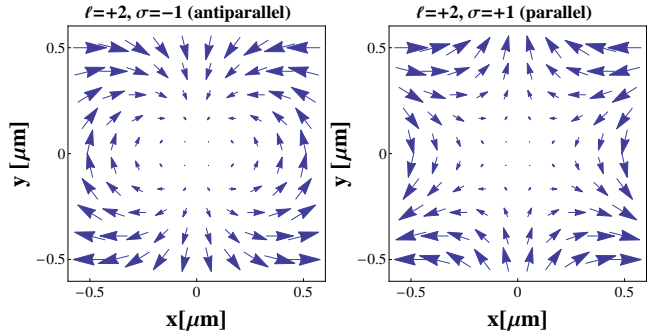}\hspace{2pc}%
\begin{minipage}[b]{14pc}\caption{\label{fig1} Electric field
patterns for the antiparallel (left) and parallel (right) class for an orbital
angular momentum of $\ell=2$.}
\end{minipage}
\end{figure}

Starting from the vector potential for a Bessel beam with given values of
$\ell$ and $\sigma$ propagating in $z$-direction, which is an exact solution
of the vectorial Helmholtz equation, the behavior of the electric and
magnetic field in a region close to the beam axis has been derived in
\cite{quinteiro2015formulation}. Remarkably, it has been found that
especially the antiparallel class is characterized by unconventional features
which become particularly strong in the case of tightly focused beams: (i)
The electric field close to the beam axis is dominated by its longitudinal
component; (ii) for $|\ell| \ge 2$ the beam in that region is dominated by
the magnetic field. The virtue of employing the longitudinal electric field
component of antiparallel beams with $|\ell|=1$ for the excitation of LH
excitons has already been discussed in \cite{quinteiro2014light}. Here we
complement this work by showing that the use of the dominant magnetic field
present in an antiparallel beam with $|\ell|=2$ again opens up new excitation
pathways which are usually unaccessible.

Close to the singularity the field amplitudes read (see Table 1 in
\cite{quinteiro2015formulation})
\begin{align}
    \tilde{E}_x(\mathbf r_\perp) \,&=\, i  \frac{\tilde{E}_0}{8}  (q_r r)^2 \, e^{i 2 \varphi}\ ,
            &\qquad \qquad &
        \tilde{B}_x(\mathbf r_\perp) \,=\, - \frac{\tilde{B}_0} {2}
            \left (\frac {q_r} {q_z} \right)^2 \ , \nonumber \\
    \tilde{E}_y(\mathbf r_\perp) \,&=\,    \frac{\tilde{E}_0}{8}  (q_r r)^2 \, e^{i 2 \varphi}\ ,  &&
        \tilde{B}_y(\mathbf r_\perp) \,=\, - i \frac {\tilde{B}_0} {2}
            \left (\frac {q_r} {q_z} \right)^2 \ , \nonumber \\
    \tilde{E}_z(\mathbf r_\perp) \,&=\, - \frac{\tilde{E}_0}{2} \frac{q_r}{q_z} (q_r r) \, e^{i \varphi}\ , &&
        \tilde{B}_z(\mathbf r_\perp) \,=\,  -i \frac {\tilde{B}_0} {2} \frac {q_r} {q_z}
            (q_r r) e^{i \varphi}\ . \nonumber
\end{align}
Here, $q_r$ and $q_z$ are the transverse and longitudinal wave vector, where
$q_r$ is inversely proportional to the beam waist, $\mathbf r_\perp$ is the
in-plane coordinate with $|\mathbf r_\perp|=r$ and azimuthal angle $\varphi$.
We will consider the excitation by tightly focused TL, i.e., beams with the
paraxial parameter $q_r/q_z \approx 1$. $\tilde{E}_0$ and
$\tilde{B}_0=\tilde{E}_0/c$ are the respective amplitudes.

When approaching the beam center, i.e. for $r\to 0$, we find that the
electric field $\mathbf E$ and the longitudinal magnetic field $B_z$ vanish
while the transverse magnetic field ${\mathbf B}_{\perp}$ remains finite.
Therefore, the transverse magnetic field dominates for interaction with
particles or nanostructures, which are much smaller than the beam waist and
are placed at or close to $r=0$, which is in fact a surprising effect at
optical frequencies.

Recently, we have developed the theory of light-nanostructure interaction for
the electric and magnetic components of general antiparallel TL beams
\cite{quinteiro2015formulation,quinteiro2017formulation}. It turned out that
the interaction can be written in a form similar to the well-known dipole
approximation for electric and magnetic fields
\cite{quinteiro2017formulation}. In particular, for TL with $\ell=2$ and
$\sigma=-1$, in which the magnetic field dominates, the Hamiltonian can be
written as \cite{quinteiro2017formulation}
\begin{equation}\label{Hamiltonian}
   H_I = - \frac{q}{2m} \mathbf B_\perp(t) \cdot (\mathbf r \times \mathbf p)
   \ =\   -i \frac{q \tilde{B}_0}{m}  \left(r_+ \, p_z - z \, p_+ \right) e^{-i \omega t} + \mathrm{c.c.}
\end{equation}
with $q$ and $m$ being the electron charge and mass, respectively, $\omega$
is the frequency of the light, $r_{\pm} =x \pm iy$,  $p_{\pm} =p_x \pm i
p_y$, and $\mathrm{c.c.}$ denoting the complex conjugate.\\[0.5ex]

\noindent\textbf{Quantum dot light-hole states}\\
We consider a semiconductor QD, where the conduction band states are $s$-type
($|s\rangle$) and can be classified by the electron spin given by the quantum
number $s_z=\pm 1/2$. For the valence band states ($|p_x\rangle,|p_y\rangle,
|p_z\rangle$), due to their $p$-type character, one finds states with total
orbital angular momenta  $j=3/2$ and $j=1/2$, where the $j=3/2$ states are
split into heavy holes with $j_z=\pm 3/2$ and LHs with $j_z=\pm 1/2$. Here,
we focus in the LH states, such that the Bloch function for electrons and
holes read:
\begin{eqnarray}
\label{Eq:LH_states}
\mid  {s_z=+1/2} \rangle \, =\, |s\rangle \mid \uparrow  \rangle &\quad \qquad &
\mid {j_z=+1/2} \rangle \, =\, \frac{1}{\sqrt{6}}\left[
\left( \mid{p_x}\rangle - i \mid{p_y}\rangle \right)  \mid \uparrow  \rangle
               + 2\mid{p_z}\rangle \mid \downarrow  \rangle \right] \nonumber \\
\mid  {s_z=-1/2} \rangle \, =\, \mid {s}\rangle \mid \downarrow  \rangle  &&
\mid  {j_z=-1/2} \rangle \, =\, -\frac{1}{\sqrt{6}} \left[ {
\left( \mid {p_x}\rangle + i \mid{p_y}\rangle \right)} \mid \downarrow  \rangle
         - 2\mid{p_z}\rangle  \mid \uparrow  \rangle  \right] \nonumber
\end{eqnarray}
where the arrow indicates the electron spin. Note that the sign of $j_z$
refers to the angular momentum of the holes which is opposite to that of
valence band electrons.

To describe the full wave function of electrons and holes in the QD, we
further apply the envelope function approximation. Within this scheme the
Bloch functions are multiplied by the respective envelope functions for
electrons and holes,
\begin{equation}
{\cal F}_{n,m}^{e/h}(\vr) = {\cal R}^{e/h}_{n,m}(r)\, e^{i m \varphi}\, {\cal Z}^{e/h}(z)\ ,
\end{equation}
which have been further separated into the radial part ${\cal
R}^{e/h}_{n,m}(r)$, an envelope angular momentum part $e^{i m \varphi}$, and
a part ${\cal Z}^{e/h}(z)$ in $z$-direction. Assuming a flat QD, we restrict
ourselves to the ground state in $z$-direction assuming wave functions ${\cal
Z}^{e/h}$ with well-defined parity.\\[0.5ex]

\noindent\textbf{The interaction of twisted light with light holes}\\
To calculate the matrix elements between LH valence and conduction band
states we follow the standard procedure in the envelope-function formalism:
We split the integration into a sum over all lattice vectors $\vR$ and an
integral over the unit cell indicated by the coordinate $\vr'$
\cite{haug2004qua}. Finally, we replace the sum over lattice vectors by an
integral over the whole system.

Following this procedure, in the Hamiltonian (\ref{Hamiltonian}) the products
of coordinates and momenta separate into terms such as $(R_+ + r_+')(P_z +
p_z')$. Keeping in mind the orthogonality and parity of the Bloch functions
under study, one finds that all matrix elements involving only intracell
coordinates $\vr'$ or only envelope coordinates $\vR$ vanish. Furthermore,
the integral over the envelope functions including either $Z$ or $P_z$
vanish, i.e., $\int dZ \, {\cal Z}^{e*}(Z) \, Z \, {\cal Z}^h(Z) = \int dZ\,
{\cal Z}^{e*}(Z) \, P_z \, {\cal Z}^h(Z)  =0$, due to the same parity of the
ground state functions ${\cal Z}^{e/h}(z)$. The only remaining matrix element
then reads
\begin{equation} \label{eq:matrixelement}
   \langle n',m'; \, s_z \mid H_I \mid n,m; \, j_z \rangle = -i \frac{q \tilde{B}_0}{m}
         \int d^3 R  \, {\cal F}_{n',m'}^{e*}(\mathbf R)  \, \left[  R_+ M_{p_z}
          -  P_+  M_{z}  \right]  {\cal F}_{n,m}^h (\mathbf R) \delta_{s_z,-j_z}\ ,
\end{equation}
where we have introduced the microscopic matrix elements
\begin{equation}
        M_{z} = \langle s_z = \pm 1/2 | \, z' \, |j_z = \mp 1/2\rangle \qquad
        M_{p_z}  = \langle s_z = \pm 1/2 | \, p_z' \, |j_z = \mp 1/2\rangle \nonumber \ .
\end{equation}
In Eq.~(\ref{eq:matrixelement}), we find that due to the optical selection
rules only electron-hole pairs having opposite angular momenta of the Bloch
state are excited, because only the matrix element $M_z$ and $M_{p_z}$
survive. In other words, only excitons with a total angular momentum of zero
(i.e. $J_z=s_z+j_z=0$) are excited. This is very different compared to the
excitation with plane waves. Plane waves have an angular momentum of $\pm 1$
and thereby excite excitons with a total angular momentum $J_z=\pm 1$. To
excite LH excitons in a QD with $J_z=0$, one would have to apply an
excitation from the side of the QD, which typically requires cleaving the
sample.

We can now evaluate the macroscopic integrals over $R$ for the two different
parts. For the first term, we rewrite $R_{+} = R e^{i\Phi}$. For the second
term we write the momentum operator in cylindrical coordinates $P_+ = -i\hbar
e^{i\Phi} \left[ \partial_R + (i/R) \partial_{\Phi}\right]$. Thus, both terms
contribute a term $\sim e^{i\Phi}$ to the angle integral. This shows, that by
exciting the QD with a magnetic TL beam having $\ell=2$ and $\sigma=-1$, the
envelope state changes by $\Delta m=1$ while the total band+spin angular
momentum of the exciton is zero. Such a transition is forbidden for
excitation with a plane wave due to dipole selection rules, and it is also
not present in the case of excitation by a TL beam with dominant electric
field in $z$-direction (i.e., an antiparallel TL beam with $|\ell|=1$)
\cite{quinteiro2014light}.

\begin{figure}[h]
\centering{\includegraphics[width=0.5\textwidth]{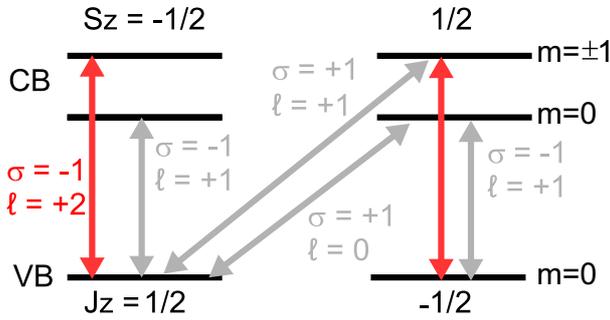}\hspace{2pc}}
\begin{minipage}[b]{14pc}\caption{\label{fig2} Optical transition between light-hole and
conduction band state in a QD induced by single pulses of twisted light having $\ell=2$
and $\sigma=-1$. For comparison also transition due to beams with $\ell=0$ and $\ell=1$
are included \cite{quinteiro2014light}. Note that for clarity reasons holes
states with $m=\pm 1$ have been removed.}
\end{minipage}
\end{figure}

Figure~\ref{fig2} combines our new results for a TL beam with $\ell=2$ and
$\sigma=-1$ with previous results reported in Ref.~\cite{quinteiro2014light}
for beams with $\ell=0$ and $\ell=1$. Note that for reasons of clarity the
Figure shows only the lowest hole shell with $m=0$. The diagonal transitions
with $\Delta m = 0$ and $\Delta m = -1$ induced correspondingly by TL beams
with ($\sigma=-1, \ell=0$) and ($\sigma=-1, \ell=-1$) are not depicted. We
thus find, that by using TL beams with $|\ell| \le 2$ all possible exciton
states in the lowest two shells with $m=0$ and $m=\pm 1$ can be excited.\\[0.5ex]

\noindent\textbf{Conclusions}\\
We have demonstrated that the optical-magnetic interaction in case of
excitation with a TL beam with OAM $\ell=2$ and SAM $\sigma=-1$ generates LH
exciton states with no band+spin angular momentum $(J_z=0)$ but with envelope
angular momentum $m=1$, and thus produces states that are not accessible with
plane waves or even TL with dominant electric interaction. Combining these
findings with previous results for beams with other values of OAM and SAM, we
find that TL beams are beneficial to obtain a full control of exciton states
in a QD.

\providecommand{\newblock}{}

\end{document}